%% file: mpm-prl.tex
\documentclass[aps,
prl,
reprint,nofootinbib,
superscriptaddress,
longbibliography
]{revtex4-2}

\usepackage[centertags]{amsmath}
\usepackage{amssymb}
\usepackage{latexsym}
\usepackage{enumerate}
\usepackage{graphicx}
\usepackage{mathrsfs}
\usepackage[hidelinks]{hyperref}
\usepackage{standalone}

\usepackage{tikz}

\usepackage{stmaryrd}
\usepackage{import}
\usepackage{tensor}
\usepackage{xcolor}
\usepackage{bm}
\usepackage{multirow}
\usepackage{breakurl}
\usepackage{float}
\usepackage{siunitx}
\usepackage{comment}
\usepackage{mathtools}
\usepackage{autobreak}
\usepackage{subfig}
\usepackage{orcidlink}
\usepackage[most]{tcolorbox} 
\allowdisplaybreaks

\usepackage[bottom]{footmisc}

\mathtoolsset{centercolon}
\usepackage{comment}

\definecolor{cadmiumgreen}{rgb}{0.0, 0.42, 0.24}

\allowdisplaybreaks

\begin{document}

\title{Analytical Time-Domain Waveforms for Black-Hole Scattering\\ in the Post-Minkowskian Expansion}

\begin{abstract}
    We present the first analytic computation of the time-domain multipoles of the gravitational waveform from hyperbolic black-hole scattering at leading order in the post-Minkowskian expansion, to all orders in velocity. We perform the partial-wave projection as a cut loop integral,
    which enables the use of dimensional regularization and associated integration techniques. As opposed to the frequency domain, the time-domain multipoles are logarithmic and do not contain Bessel-like functions.
    Our results pave the way for the incorporation of post-Minkowskian fluxes in waveform models for binary mergers.
\end{abstract}

\author{Stefano De Angelis}
\affiliation{Institut de Physique Th\'eorique, CNRS, CEA, Universit\'e Paris-Saclay, 91191 Gif-sur-Yvette cedex, France}

\author{Vincent F. He}
\affiliation{Institut des Hautes \'Etudes Scientifiques, 91440 Bures-sur-Yvette, France}

\author{Julio Parra-Martinez}
\affiliation{Institut des Hautes \'Etudes Scientifiques, 91440 Bures-sur-Yvette, France}

\maketitle


\noindent\textbf{Introduction.} Gravitational-wave observations by the LIGO-Virgo-KAGRA
collaboration~\cite{LIGOScientific:2016aoc,KAGRA:2021vkt,LIGOScientific:2025slb} are interpreted
with waveform models that combine numerical relativity with analytic
approximations to the relativistic two-body problem. 
Widely used waveform families include effective-one-body (EOB) models~\cite{Buonanno:1998gg,Buonanno:2000ef,Pompili:2023tna,Nagar:2023zxh,Gamboa:2026jht}, phenomenological models~\cite{Ajith:2007qp,Hamilton:2025xru}, and numerical-relativity (NR) surrogate models~\cite{Field:2013cfa,Ravishankar:2026akf}.
A complete description requires understanding both the motion of the
bodies and the gravitational radiation they emit. In EOB models, the
analytic input comprises a Hamiltonian, which describes
the conservative dynamics, and the multipolar waveform modes $f_{\ell m}$,
which determine the radiation-reaction force through the fluxes of
energy and angular momentum~\cite{Damour:2008gu}.

Traditionally, these ingredients were mainly computed in the
post-Newtonian (PN) expansion~\cite{Blanchet:2013haa,Blanchet:1985sp,Thorne:1980ru}, which assumes small velocities and weak
fields. In recent years, the post-Minkowskian (PM) expansion, which assumes
weak fields but arbitrary velocities, has been pushed to high precision for
black-hole scattering observables. The scattering angle and impulse are
known through fourth order in $G$~\cite{Bern:2019nnu,Damour:2020tta,DiVecchia:2021ndb,Herrmann:2021lqe,DiVecchia:2021bdo,Herrmann:2021tct,Bern:2021dqo,Bern:2021yeh,Dlapa:2021npj,Dlapa:2022lmu,Jakobsen:2023ndj,Jakobsen:2023hig}, with partial results available at fifth order~\cite{Driesse:2024xad,Driesse:2024feo,Bern:2025wyd,Driesse:2026qiz}. Some of these results have been
incorporated into EOB models, mainly through the
Hamiltonian~\cite{Damour:2016gwp,Damour:2017zjx,Antonelli:2019ytb,Khalil:2022ylj,Buonanno:2024vkx,Buonanno:2024byg,Damour:2025uka}
and, for scattering orbits, through the radiation-reacted
angle~\cite{Damour:2022ybd,Rettegno:2023ghr}.
The gravitational waveform is known at
leading~\cite{Kovacs:1977uw,Kovacs:1978eu,Jakobsen:2021smu,Mougiakakos:2021ckm,DeAngelis:2023lvf,DeAngelis:2026ymn,Jakobsen:2021lvp}
and next-to-leading
order~\cite{Brandhuber:2023hhy,Herderschee:2023fxh,Elkhidir:2023dco,Georgoudis:2023lgf,Bini:2023fiz,Bohnenblust:2023qmy,Bohnenblust:2025gir},
but has not yet been used directly as input to EOB waveform models, whose
modes and radiation reaction are still built from resummed PN results. 
Obtaining analytic time-domain multipoles from the available
direction-dependent waveforms requires an angular projection,
together with a Fourier transform when starting in the frequency
domain. Moreover, matching amplitude-based PM waveforms to PN multipoles
requires consistent and subtle choices of frame~\cite{Georgoudis:2023eke,Bini:2024rsy,Georgoudis:2024pdz,Bini:2023fiz,Elkhidir:2024izo}.

In this Letter we initiate a program to compute, analytically and to all
orders in velocity, the time-domain PM multipoles 
of the gravitational waveform emitted in the scattering of two
nonspinning black holes. 
We write the
projection onto spin-weighted spherical harmonics as an integral over the
on-shell graviton momentum, with the harmonics expressed as covariant
symmetric-trace-free tensors.
Combined with the Fourier
transform to impact-parameter space and the integral over the graviton
frequency, the projection becomes a cut Feynman integral with
non-integer propagator powers. This form allows the use of reverse
unitarity~\cite{Anastasiou:2002yz,Anastasiou:2002qz},
integration-by-parts reduction~\cite{Tkachov:1981wb,Chetyrkin:1981qh,Laporta:2000dsw} and canonical differential equations~\cite{Kotikov:1990kg,Remiddi:1997ny,Gehrmann:1999as,Henn:2013pwa,Parra-Martinez:2020dzs}.
The time-domain multipoles involve elementary functions familiar from Feynman integrals, whereas frequency-domain waveform results involve Bessel-like functions~\cite{Brunello:2024ibk,Bini:2024ijq,Brunello:2025eso}.
A similar strategy was recently applied to the computation of the leading nonlinear memory~\cite{Georgoudis:2025vkk}.

Integrating over the emission angles yields individual waveform modes
with fewer kinematic variables, providing a natural starting point for
mode-by-mode resummation~\cite{Damour:2008gu,Ivanov:2025ozg} and for investigating
analytic continuation from scattering to bound waveforms \cite{Kalin:2019rwq,Kalin:2019inp,Cho:2021arx,Dlapa:2024cje,Bini:2024tft,Dlapa:2025biy,Dlapa:2026oyq}.
The modes also provide a common basis for comparing PM calculations
based on scattering amplitudes with PN
~\cite{Bini:2024ijq,Bini:2026dvn,Bini:2026jwc},
NR and self-force calculations~\cite{Long:2021ufh,Fucito:2024wlg,Long:2025tvk,Warburton:2025ymy,Barack:2023oqp,Geralico:2026jit}.

As a first application, we present the first closed analytic expressions
for the time-domain quadrupolar and octupolar modes of the leading-order
PM waveform at generic velocity, in the rest frame of one of the black
holes. The expressions retain the full dependence on velocity and
retarded time and involve only rational functions, logarithms. Their unexpanded velocity dependence encodes an
infinite sequence of PN corrections at fixed PM order, providing
analytic benchmarks beyond any finite velocity truncation.
We validate the results against direct numerical projection of the
waveform, an independent self-force calculation \cite{WaveformSF}, and the PN expansion
of the multipolar post-Minkowskian (MPM) waveform, finding full agreement.
These results establish explicit
relativistic counterparts of the PN modes used in waveform models
and demonstrate a systematic route towards 
higher multipoles,
higher PM orders, and the incorporation of radiative PM information
into waveform modeling.

\begin{figure}[t!]
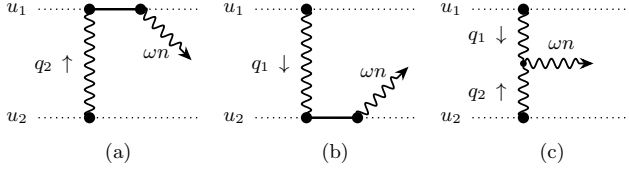

\centering
\includestandalone[width= \columnwidth]{wqft_tree}
\caption{WQFT diagrams contributing to the leading-order waveform.}
\label{fig:wqft}
\end{figure}

\noindent \textbf{Time-domain waveform.} In the PM expansion, metric perturbations are expanded around flat space as $g_{\mu\nu} = \eta_{\mu\nu} + h_{\mu\nu}$. The gravitational waveform is defined as the coefficient of the radiative field at future null infinity:
\begin{equation}
\varepsilon^{\mu \nu} h_{\mu\nu}( u t^\alpha+rn^\alpha) \sim 
\frac{4 G}{r^{\frac{D-2}{2}}} f(u,n)\,.
\end{equation}
Here $\varepsilon$ is the polarization tensor, $u$ is retarded time, $t^\alpha$ is a unit time-like vector, and $n^\alpha$ is a null vector normalized as $n\cdot t=1$. By a standard saddle-point argument at large $r$
~\cite{Cristofoli:2021vyo}, the waveform is related to the
coefficient of the plane-wave mode $e^{-i\omega n\cdot x}$,
which we denote by $W(\omega n)$, through
\begin{equation}
    f(u,n)=\frac{-i} {8G}\int_0^\infty \frac{d\omega}{(2\pi )^{\frac{D}{2}}}(-i\omega)^{\frac{D-4}{2}} e^{-i\omega u} W(\omega n)+\text{c.c.}
    \label{waveform}
\end{equation}
In binary black hole scattering, $W(\omega n)$ can be represented as the
Fourier transform of an amplitude-like object $\mathcal{R}$
into impact-parameter space~\cite{Kosower:2018adc,Cristofoli:2021vyo}. Namely,
\begin{align}
    W(\omega n)= \!\int\! \prod_{j=1,2}&\frac{d^D q_j}{(2\pi)^{D-1}} 
    \delta(q_j\cdot u_j) e^{i q_j\cdot b_j} \\
    &\times(2\pi)^D \delta^{(D)}(\sum_jq_j-\omega n) \mathcal{R}(\varepsilon, n ,q_j) . \nonumber
\end{align}
Here $q_j$, $u_j$, and $b_j$ are the momentum transfer, four-velocity, and impact parameter of black hole $j$, respectively.
$\mathcal{R}$ is related to the classical
limit of five-point amplitudes and their 
cuts~\cite{Cristofoli:2021vyo,Caron-Huot:2023vxl,Caron-Huot:2023ikn,Biswas:2024ept,Georgoudis:2023eke,Brunello:2024ibk}.
At leading PM order, $\mathcal{R}$ is simply the tree-level five-point amplitude, which is a rational function. To obtain the leading-order classical amplitude $W$, we use generalized unitarity in worldline quantum field theory (WQFT)~\cite{Mogull:2020sak,He:2025how,Jakobsen:2022psy,Haddad:2025cmw}. The corresponding Feynman diagrams are shown in Fig.~\ref{fig:wqft}. 

At each PM order, if one redefines $q_i = \omega \tilde{q}_i$, then $\mathcal{R}$ scales homogeneously in $\omega$. Thus, the energy integration in Eq.~\eqref{waveform} can be performed with the following result:
\begin{equation}
 \label{eq:time_domain_ips}
 \begin{split}
f(u,n)=\frac{1}{8G\,(2\pi)^{\frac{D}{2}}}\int\frac{d^D\tilde q}{(2\pi)^{D-2}}\,
\delta(\tilde q\cdot u_1)\,\delta((n-\tilde q)\cdot u_2)\\
\times\left(\frac{(-i)^{\frac{D-2}{2}+\nu}\,\Gamma(\nu)\,\mathcal R(\varepsilon,n,\tilde q)}
{\big[u-(\tilde q\cdot b_1)-(n-\tilde q)\cdot b_2-i\epsilon\big]^{\nu}}+\text{c.c.}\right),
\end{split}   
\end{equation}
where momentum conservation has been used to set $\tilde q_1=\tilde q$ and $\tilde q_2=n-\tilde q$, and the exponent $\nu$ is fixed by dimensional analysis and depends on the PM order:
\begin{equation}
    \nu = \frac{3}{2}D-5+(N-1)(D-3)
\end{equation} 
in $D$ dimensions at $N$th PM order. 

\noindent \textbf{Covariant partial waves and reverse unitarity.} It is natural to consider the multipolar decomposition of the waveform,
\begin{equation}
     f(u,n) = \sum_{\ell\ge 2} \sum_{|m|\le \ell} f_{\ell m}(u) \,\,{}_hY_{\ell m}(n)\,,
\end{equation}
where ${}_hY_{\ell m}$ are spin-weighted spherical harmonics with helicity $h=\pm2$. In a frame defined by the 
time-like vector $t^\mu$, and an orthonormal space-like triad
$\{x^\mu,y^\mu,z^\mu\}$, these can be represented by
symmetric-trace-free (STF) tensors as
\begin{align}
  &{}_{h}Y_{\ell m}
  = {}_hN_{\ell m} \,  Y_{\mu_1\cdots \mu_\ell}\,
  e_h^{\mu_1} e_h^{\mu_2}\,\hat n^{\mu_3}\cdots\hat n^{\mu_\ell},
\end{align}
where 
\begin{equation}
  Y_{\mu_1\cdots \mu_\ell}  =  z_{\langle\mu_1}\cdots z_{\mu_{\ell-|m|}}\,
  \mathfrak{m}_{\mu_{\ell-|m|+1}}\cdots\mathfrak{m}_{\mu_\ell\rangle}\,,
\end{equation}
with $\mathfrak{m}^\mu = x^\mu + i\,\mathrm{sgn}(m)\,y^\mu$, and 
\begin{equation}
\label{eq:norm}
   {}_hN_{\ell m}=\sqrt{\tfrac{2^{|h|}\,(\ell!)^{2}\,[(2\ell+1)!!]^{2}}
             {4\pi\,(2\ell+1)\,(\ell+m)!\,(\ell-m)!\,(\ell-|h|)!\,(\ell+|h|)!}}.
\end{equation}
Here $\hat n^\mu$ and $e^\mu$ are the unit spatial direction pointing inward and transverse
polarization in the rest frame of $t^\mu$, expressed covariantly in terms of
the graviton momentum $n^\mu$, normalized to $t\cdot n=1$, and polarization
$\varepsilon^\mu$:
\begin{equation}
  \hat n^\mu = t^\mu - n^\mu\,, \quad
  e^\mu = \varepsilon^\mu - n^\mu\,(t\cdot\varepsilon)\,.
\end{equation}
Note that with these definitions, ${}_{h}Y_{\ell m}$ is manifestly gauge invariant.  

The partial-wave projection onto linearly
polarized waveform modes can be performed as
\begin{widetext}
\begin{align}
    \label{eq:fplus}
   f^{+}_{\ell m}(u)
&=
\frac{1}{2}
\int d^D n\,\delta(n^2)\delta(t\cdot n-1)
\sum_{h=\pm 2} f(u,n)\,{}_hY_{\ell m}(n)
+\text{c.c.},\\
   \label{eq:fcross}
     f^{\times}_{\ell m}(u)
&=
\frac{1}{2i}
\int d^D n\,\delta(n^2)\delta(t\cdot n-1)
\sum_{h=\pm 2} f(u,n)\,\frac{1}{2}\mathcal{H}\,{}_hY_{\ell m}(n)
+\text{c.c.},
\end{align}
\end{widetext}
where the helicity operator is defined as
\begin{equation}
    {\cal H} = \frac{t\cdot \mathcal{P}}{t\cdot k}\,,
\end{equation}
where $\mathcal{P}$ is the Pauli--Lubanski vector,
\begin{equation}
   \mathcal{P}^\mu = \frac12\, \epsilon^{\mu\nu\sigma\rho} k_\nu J_{\sigma\rho}\,,\quad J^{\sigma\rho} = i k^{[\sigma}\partial_k^{\rho]} + i \varepsilon^{[\sigma}\partial_\varepsilon^{\rho]}\,.
\end{equation}
Acting on the polarization dependence of the harmonics, the helicity
operator $\mathcal H$ has eigenvalues $h=\pm2$. The sum over the two physical polarizations is performed with the
four-dimensional completeness relation
\begin{equation}
\sum_{\text{pol}}\varepsilon^{*\mu}\varepsilon^{*\nu}\varepsilon^\alpha\varepsilon^\beta
=\frac{1}{2}\left(\eta^{\mu\alpha}\eta^{\nu\beta}+\eta^{\mu\beta}\eta^{\nu\alpha}-\eta^{\mu\nu}\eta^{\alpha\beta}\right),
\end{equation}
valid provided both $f$ and ${}_hY_{\ell m}$ are gauge invariant.

This representation allows us to compute the partial-wave coefficients, together with the Fourier transform to the time domain and impact-parameter space in Eq.~\eqref{eq:time_domain_ips}, as cut Feynman integrals with non-integer propagator powers. Besides being well suited to mode-by-mode EOB waveform
modeling~\cite{Damour:2008gu,Damour:2016gwp}, the partial-wave projection simplifies the computation by trading the two emission angles for one additional loop
integration and avoids the Bessel-like functions that appear in the frequency domain (see, e.g., \cite{Brunello:2024ibk,Brunello:2025eso}).

Furthermore, in this work we choose to work in the rest frame of the first black hole, i.e. $t^\mu=u_1^\mu$ and $b_1^\mu=0$. We also orient our frame such that the spatial component of $u_2^\mu$ is in the $x$-direction and $b^\mu = b_2^\mu - b_1^\mu$ points in the $z$-direction. The orthonormal vectors are:
\begin{equation}
    x^\mu=\frac{u_2^\mu-\gamma u_1^\mu}{\sqrt{\gamma^2-1}}, \quad z^\mu=\frac{b^\mu}{|b|},
\end{equation}
where $\gamma=u_1\cdot u_2$. The remaining unit vector $y^\mu$ is fixed up to a sign by its orthogonality to $t^\mu$, $x^\mu$, and $z^\mu$.

With these conventions, the spherical harmonics are simple functions of Lorentz-invariant products of kinematics, which can be simply treated as a numerator for the Feynman integrals. For example,
\begin{align}
 {}_hY_{20}
    &=\frac{1}{2|b|^2} \sqrt{\frac{15}{2 \pi }}\, \big[(b\cdot\varepsilon)-(n\cdot b)(u_1\cdot\varepsilon)\big]^2.
\end{align}
More generally, an $\ell$-th multipole yields a rank-$\ell$ numerator in the integrated momenta.

Given this representation of the waveform multipoles, for any value of $(\ell,m)$, we can use integration-by-parts (IBP) identities \cite{Tkachov:1981wb,Chetyrkin:1981qh,Laporta:2000dsw} and the reverse-unitarity method \cite{Anastasiou:2002yz,Anastasiou:2002qz,Herrmann:2021lqe} to reduce them to a basis of master integrals. Using
LiteRed2~\cite{Lee:2012cn,Lee:2013mka} we obtain a basis of eight master
integrals, and derive their differential
equations~\cite{Kotikov:1990kg,Remiddi:1997ny,Gehrmann:1999as,Parra-Martinez:2020dzs} with respect
to the two variables $t$ and $x$ which are related to the retarded time $u$ and the relative Lorentz factor $\gamma$ by
\begin{equation}
    \gamma=\frac{1}{2}\left(x+\frac{1}{x}\right)\,, \quad
    \frac{u}{|b|}=\frac{x(t^2-1)}{t(1-x^2)}\,.
\end{equation}
We choose the physical branch $0<x<1$ and $t>0$.
Here the scalar variables $x, t$ are distinct from the triad vector $x^\mu$ and the observer vector $t^\mu$.
The dependence on the remaining scale $|b|$ can be recovered by dimensional analysis.
We then bring the differential equations to canonical
form~\cite{Henn:2013pwa}, compute the static boundary conditions in the low-velocity limit ($x\to1$), and solve the master integrals by path-ordered iterated integration.
A more detailed exposition of the integrals will appear elsewhere.
 
\begin{figure*}[tbp!]
  \centering
\includegraphics[width=0.48\textwidth]{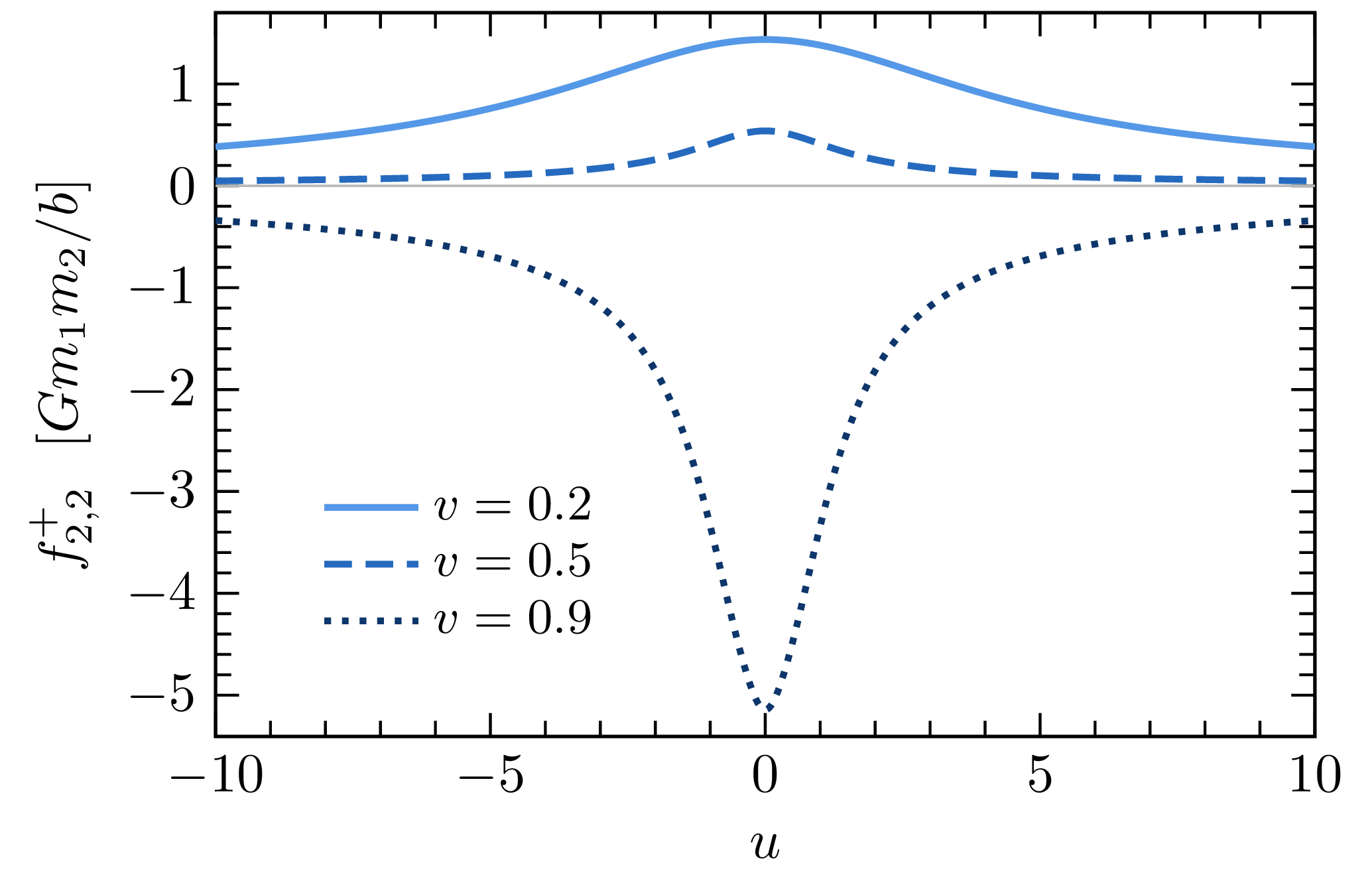}
  \hfill
  \includegraphics[width=0.48\textwidth]{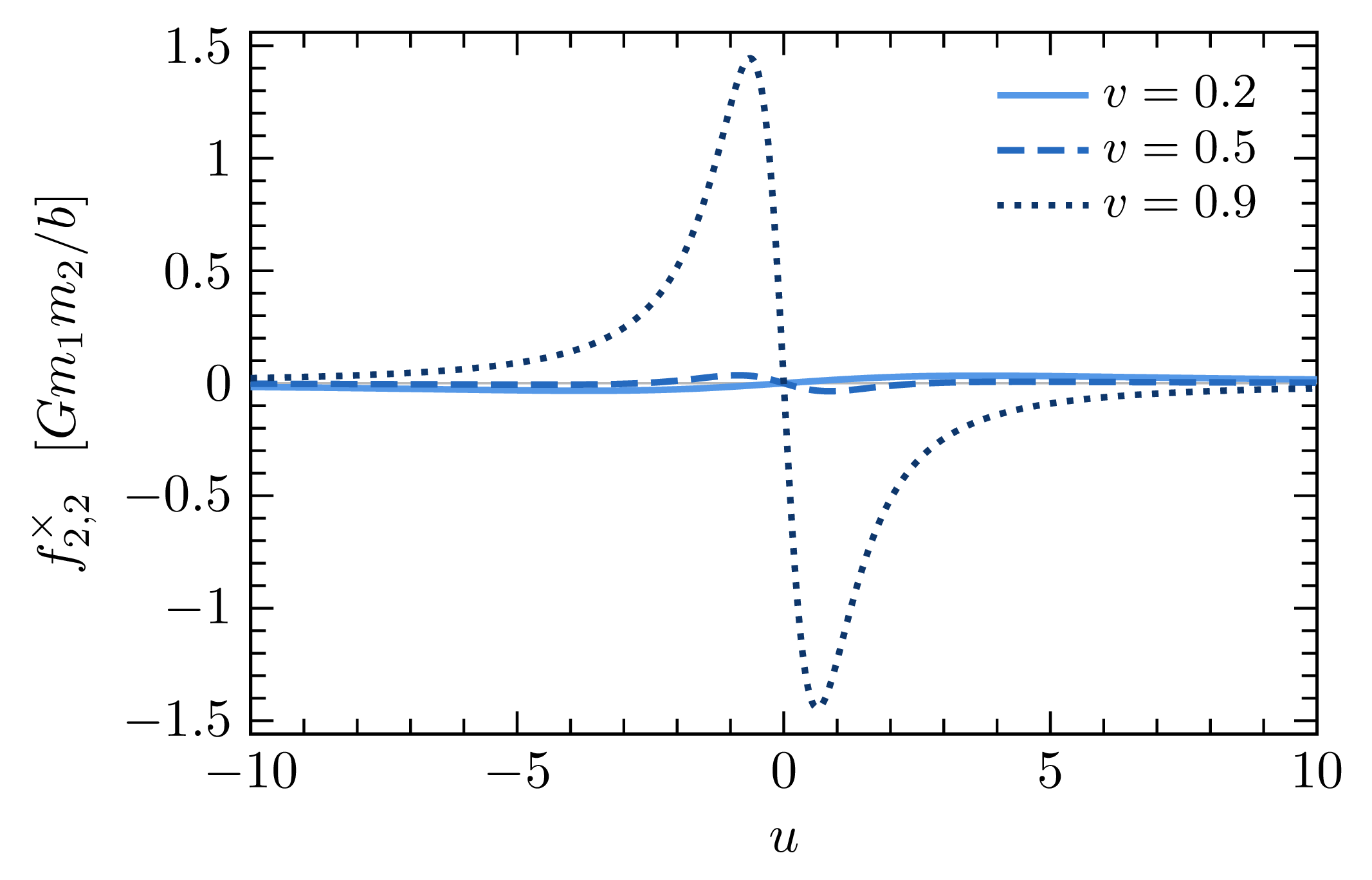}
  \caption{The quadrupolar waveform multipoles  $f^{+}_{22}$ (left) and $f^{\times}_{22}$ (right) as
  functions of the retarded time, $u$, for various values of the velocity $v$, where $v=(1-x^2)/(1+x^2)$.}
  \label{fig:f22}
\end{figure*}

\noindent \textbf{Results.} The leading-PM multipole moments take the form
\begin{widetext}
        \begin{equation}
\begin{split}
f^{\mathfrak{p}}_{\ell m}(x,t) = \frac{8\pi G\, m_1m_2\,{}_2N_{\ell m}}{|b|\sqrt{\gamma^2-1}}\frac{1}{(\gamma^2-1)^{\ell/2}}
\Bigg[& {}_0R^\mathfrak{p}_{\ell m}
+ {}_1R^\mathfrak{p}_{\ell m}\,\log x
+ {}_2R^\mathfrak{p}_{\ell m}\,\arctan\!\left(\frac{t(1-x^2)}{x(1+t^2)}\right) \\
&+ {}_3R^\mathfrak{p}_{\ell m}\,\log\!\left(\frac{(1+t^2)\,x}{1+t^2x^2}\right)
+ {}_4R^\mathfrak{p}_{\ell m}\,\log\!\left(\frac{(1+t^2)\,x}{t^2+x^2}\right) \Bigg],
\end{split}
\end{equation}
    \end{widetext} 
where the coefficients ${}_iR^{\mathfrak p}_{\ell m}$ are rational functions of $x$ and $t$, and $\mathfrak p\in\{+,\times\}$ labels the polarization.
The overall poles at $\gamma=1$ ($x=1$) are spurious. 
The last two logarithms are exchanged under either $t\to1/t$ or $x\to1/x$. Correspondingly, the coefficients are related by
\begin{equation}
{}_4R^{\mathfrak{p}}_{\ell m}(t,x)= \pm(-1)^{\ell-m}\,{}_3R^{\mathfrak{p}}_{\ell m}(1/t,\,x)= \mp\,{}_3R^{\mathfrak{p}}_{\ell m}(t,\,1/x),
\label{eq:R4R3}
\end{equation}
where the upper (lower) sign applies to the $+$ ($\times$) polarization.

Additional logarithms appear at intermediate stages of the calculation, but we have verified through $\ell=4$ that the final multipole moments contain only the four transcendental functions displayed above. We expect this space of functions to suffice for all $(\ell,m)$.

Since the scattering takes place in the $xz$-plane, the system is symmetric under
$y\to-y$. This reflection sends $m\to-m$ and exchanges the two graviton
helicities. In terms of the linear polarizations,
$f^{+}$ is even under this reflection, while
$f^{\times}$ is odd. Therefore
\begin{equation}
  f^{+}_{\ell\,-m}=f^{+}_{\ell m},\quad f^{\times}_{\ell\,-m}=-f^{\times}_{\ell m},
\end{equation}
and it suffices to consider $m\ge0$. In particular, $f^{\times}_{\ell 0}=0$.

For $\ell=2$, $m=0$, the rational coefficients  are given in Table~\ref{tab:R20}; ${}_4R^{+}_{20}$ follows from Eq.~\eqref{eq:R4R3}.
\begin{table}[t]
\centering
\setlength{\fboxsep}{6pt}
\fbox{$\displaystyle
\begin{aligned}
{}_0R^{+}_{2,0} &= \frac{1}{1536\,\bar t_1x_1(\bar t_1^{2}+x_1^{2})^{4}}\Big[\bar t_1^{10}(645x_1^{4}+552x_1^{2}+48) \\
&\quad +\bar t_1^{8}(2576x_1^{6}+1408x_1^{4}-816x_1^{2}-192) \\
&\quad +\bar t_1^{6}x_1^{2}(3998x_1^{6}+368x_1^{4}-3600x_1^{2}-768)\\
&\quad +\bar t_1^{4}x_1^{4}(2720x_1^{6}-2176x_1^{4}-5520x_1^{2}-1152) \\
&\quad +\bar t_1^{2}x_1^{6}(749x_1^{6}-2232x_1^{4}-3616x_1^{2}-768) \\
&\quad -x_1^{8}(736x_1^{4}+928x_1^{2}+192)\Big] \\[6pt]
{}_1R^{+}_{2,0} &= \frac{\bar x_1(2x_1^{2}-1)[\bar t_1^{2}(2x_1^{2}+1)-x_1^{2}-1]}{8\,\bar t_1x_1^{2}}\\[6pt]
{}_2R^{+}_{2,0} &= \frac{\bar t_1^{2}(215x_1^{4}+184x_1^{2}+16)-15x_1^{6}-248x_1^{4}-416x_1^{2}-64}{512\,x_1^{2}} \\[6pt]
{}_3R^{+}_{2,0} &= \frac{4x_1^{3}[\bar t_1^{2}(2x_1^{2}+1)-x_1^{2}-1]+t_1\bar t_1\bar x_1(4x_1^{4}-1)}{16\,\bar t_1x_1^{2}}
\end{aligned}$}
\caption{Rational coefficients of the multipole $f^{+}_{20}$, with
$x_1\equiv(x-x^{-1})/2$, $\bar x_1\equiv(x+x^{-1})/2$, $t_1\equiv(t-t^{-1})/2$ and $\bar t_1\equiv(t+t^{-1})/2$.}
\label{tab:R20}
\end{table}
We include all the coefficients up to $\ell=3$ in the Supplemental Material and in the ancillary file for the arXiv version of this paper. Figure~\ref{fig:f22} shows $f^{+}_{22}$ and $f^{\times}_{22}$ as functions of the retarded time for three velocities.

We have validated our results up to $\ell=3$ against numerical integration of the leading-PM
waveform~\cite{Kovacs:1977uw,Kovacs:1978eu,Jakobsen:2021smu}, the post-Newtonian expansion of the
MPM waveform~\cite{Bini:2024rsy,Bini:2026dvn,Bini:2026jwc}, and an
independent self-force calculation of the $\ell=2$ multipole up to 20PN presented in concurrent
work~\cite{WaveformSF}, finding
complete agreement in all cases. 

\textbf{Conclusions.} We have obtained the first analytic expressions for the time-domain
multipoles of the gravitational waveform from black-hole scattering at
leading PM order and to all orders in velocity. In the rest frame of one black hole, each multipole depends on
only two variables, the velocity and the retarded time in units of the
impact parameter, and involves only rational functions and logarithms. These multipoles provide leading-PM contributions to
unbound-orbit waveform modes and their associated fluxes at arbitrary
velocity.

In future work we will present a more detailed exposition of our methods, as well as the computation of the next-to-leading-order PM multipoles. A further natural extension of our work would be to repeat the computation in the center-of-mass frame. In this frame the mass ratio enters as an additional variable in the integrals, which makes the calculation significantly more challenging.

\textbf{Acknowledgements.}
We thank Giacomo Brunello, Thibault Damour, David Kosower and Davide Usseglio for discussions.
The work of J.P.-M. is supported by the European Research
Council (ERC) under the European Union's Horizon 2020 research and innovation programme, grant agreement ERC-StG-101221094 GravitaS. The work of V.F.H. is supported by the Simons Foundation. We acknowledge the use of AI assistants
for figure generation, copy-editing, and proofreading.

\bibliographystyle{apsrev4-2-prl} 
\bibliography{bibliography}
\appendix
\onecolumngrid
\section*{Supplemental Material}
\setlength{\parskip}{0.3em}
\section{Multipole coefficients}
\input{appendix_multipoles}


\end{document}

%% file: wqft_tree.tex
\usetikzlibrary{decorations.pathmorphing,arrows.meta}
\tikzset{
  wl/.style={dotted, line width=0.6pt},
  zprop/.style={line width=1.1pt},
  grav/.style={decorate, decoration={snake, amplitude=0.75mm, segment length=1.9mm}, line width=0.8pt},
  gravout/.style={decorate, decoration={snake, amplitude=0.75mm, segment length=1.9mm, post length=1.8mm},
                  line width=0.8pt, -{Stealth[length=2.2mm, width=1.8mm]}},
  dotv/.style={circle, fill=black, inner sep=0pt, minimum size=5pt},
  sdot/.style={circle, fill=black, inner sep=0pt, minimum size=3pt},
}
\providecommand{\worldlines}{%
  \draw[wl] (-0.8,1.7) -- (1.7,1.7); \node[anchor=east] at (-0.85,1.7) {$u_1$};
  \draw[wl] (-0.8,0)   -- (1.7,0);   \node[anchor=east] at (-0.85,0)   {$u_2$};
}

\begin{tikzpicture}

\begin{scope}[shift={(0,0)}]
  \worldlines
  \draw[grav] (0,0) -- (0,1.7);
  \node[anchor=east] at (-0.15,0.85) {$q_2\,\uparrow$};
  \draw[zprop] (0,1.7) -- (0.8,1.7);
  \draw[gravout] (0.8,1.7) -- (1.6,0.9);
  \node at (1.05,1.0) {$\omega n$};
  \node[dotv] at (0,1.7) {}; \node[dotv] at (0,0) {}; \node[dotv] at (0.8,1.7) {};
  \node at (0.45,-0.55) {(a)};
\end{scope}

\begin{scope}[shift={(3.4,0)}]
  \worldlines
  \draw[grav] (0,1.7) -- (0,0);
  \node[anchor=east] at (-0.15,0.85) {$q_1\,\downarrow$};
  \draw[zprop] (0,0) -- (0.8,0);
  \draw[gravout] (0.8,0) -- (1.6,0.8);
  \node at (1.05,0.7) {$\omega n$};
  \node[dotv] at (0,1.7) {}; \node[dotv] at (0,0) {}; \node[dotv] at (0.8,0) {};
  \node at (0.45,-0.55) {(b)};
\end{scope}

\begin{scope}[shift={(6.8,0)}]
  \worldlines
  \draw[grav] (0,1.7) -- (0,0.85);
  \draw[grav] (0,0) -- (0,0.85);
  \node[anchor=east] at (-0.15,1.3) {$q_1\,\downarrow$};
  \node[anchor=east] at (-0.15,0.4) {$q_2\,\uparrow$};
  \draw[gravout] (0,0.85) -- (1.1,0.85);
  \node at (0.6,1.15) {$\omega n$};
  \node[dotv] at (0,1.7) {}; \node[dotv] at (0,0) {}; \node[sdot] at (0,0.85) {};
  \node at (0.45,-0.55) {(c)};
\end{scope}

\end{tikzpicture}

%% file: appendix_multipoles.tex
\begin{tcolorbox}[breakable, colback=white, colframe=black!60, boxrule=0.5pt, arc=2pt,
  left=4pt, right=4pt, top=2pt, bottom=2pt, title={$\ell=2,\ m=1$}, coltitle=black, colbacktitle=black!8]
\begin{align}
{}_0R^{+}_{2,1} &= -\frac{t_1\bar x_1}{768\,\bar t_1x_1(\bar t_1^{2}+x_1^{2})^{4}}\Big[\bar t_1^{8}(3x_1^{4}+408x_1^{2}+96)+\bar t_1^{6}x_1^{2}(11x_1^{4}+1560x_1^{2}+384)+\bar t_1^{4}x_1^{4}(53x_1^{4}+2280x_1^{2}+576) \nonumber\\
&\quad -\bar t_1^{2}x_1^{6}(3x_1^{4}-1448x_1^{2}-384)+x_1^{8}(368x_1^{2}+96)\Big], \label{eq:R0_21p}\\[6pt]
{}_1R^{+}_{2,1} &= -\frac{t_1\bar x_1^{2}(2x_1^{2}-1)}{8\,\bar t_1x_1^{2}}, \label{eq:R1_21p}\\[6pt]
{}_2R^{+}_{2,1} &= -\frac{t_1\bar x_1(65x_1^{4}+72x_1^{2}-16)}{256\,x_1^{2}}, \label{eq:R2_21p}\\[6pt]
{}_3R^{+}_{2,1} &= -\frac{8t_1x_1^{3}\bar x_1+\bar t_1(4x_1^{4}+8x_1^{2}+1)}{32\,\bar t_1x_1^{2}}, \label{eq:R3_21p}\\[6pt]
{}_0R^{\times}_{2,1} &= -\frac{\bar x_1}{96\,\bar t_1(\bar t_1^{2}+x_1^{2})^{3}}\Big[\bar t_1^{8}(45x_1^{2}+18)+\bar t_1^{6}(144x_1^{4}+6x_1^{2}-12)+\bar t_1^{4}x_1^{2}(147x_1^{4}-94x_1^{2}-36) \nonumber\\
&\quad +\bar t_1^{2}x_1^{4}(56x_1^{4}-126x_1^{2}-36)-x_1^{6}(52x_1^{2}+12)\Big], \label{eq:R0_21c}\\[6pt]
{}_1R^{\times}_{2,1} &= -\frac{\bar x_1^{2}(2x_1^{2}-1)(2\bar t_1^{2}-1)}{8\,\bar t_1x_1}, \label{eq:R1_21c}\\[6pt]
{}_2R^{\times}_{2,1} &= -\frac{\bar x_1[\bar t_1^{2}(15x_1^{2}+6)-10x_1^{2}-8]}{32\,x_1}, \label{eq:R2_21c}\\[6pt]
{}_3R^{\times}_{2,1} &= -\frac{2x_1^{3}\bar x_1(2\bar t_1^{2}-1)+t_1\bar t_1\bar x_1^{2}(2x_1^{2}-1)}{8\,\bar t_1x_1}. \label{eq:R3_21c}
\end{align}
\end{tcolorbox}

\begin{tcolorbox}[breakable, colback=white, colframe=black!60, boxrule=0.5pt, arc=2pt,
  left=4pt, right=4pt, top=2pt, bottom=2pt, title={$\ell=2,\ m=2$}, coltitle=black, colbacktitle=black!8]
\begin{align}
{}_0R^{+}_{2,2} &= \frac{1}{1536\,\bar t_1x_1(\bar t_1^{2}+x_1^{2})^{4}}\Big[\bar t_1^{10}(855x_1^{4}+792x_1^{2}+144)+\bar t_1^{8}(3408x_1^{6}+2712x_1^{4}+1200x_1^{2}+192) \nonumber\\
&\quad +\bar t_1^{6}x_1^{2}(5418x_1^{6}+2920x_1^{4}+3216x_1^{2}+768)+\bar t_1^{4}x_1^{4}(4160x_1^{6}+296x_1^{4}+3984x_1^{2}+1152) \nonumber\\
&\quad +\bar t_1^{2}x_1^{6}(1199x_1^{6}-1632x_1^{4}+2272x_1^{2}+768)-x_1^{8}(736x_1^{4}-544x_1^{2}-192)\Big], \label{eq:R0_22p}\\[6pt]
{}_1R^{+}_{2,2} &= \frac{\bar x_1(2x_1^{2}-1)[\bar t_1^{2}(2x_1^{2}+1)-x_1^{2}+1]}{8\,\bar t_1x_1^{2}}, \label{eq:R1_22p}\\[6pt]
{}_2R^{+}_{2,2} &= \frac{\bar t_1^{2}(285x_1^{4}+264x_1^{2}+48)-165x_1^{6}+64x_1^{4}+352x_1^{2}-64}{512\,x_1^{2}}, \label{eq:R2_22p}\\[6pt]
{}_3R^{+}_{2,2} &= \frac{4x_1^{3}[\bar t_1^{2}(2x_1^{2}+1)-x_1^{2}+1]+t_1\bar t_1\bar x_1(2x_1^{2}-1)(2x_1^{2}+1)}{16\,\bar t_1x_1^{2}}, \label{eq:R3_22p}\\[6pt]
{}_0R^{\times}_{2,2} &= \frac{t_1}{48\,\bar t_1(\bar t_1^{2}+x_1^{2})^{3}}\Big[\bar t_1^{6}(15x_1^{4}+48x_1^{2}+12)+\bar t_1^{4}x_1^{2}(32x_1^{4}+148x_1^{2}+36)+\bar t_1^{2}x_1^{4}(9x_1^{4}+144x_1^{2}+36) \nonumber\\
&\quad +x_1^{6}(52x_1^{2}+12)\Big], \label{eq:R0_22c}\\[6pt]
{}_1R^{\times}_{2,2} &= \frac{t_1\bar x_1(2x_1^{2}-1)}{4\,\bar t_1x_1}, \label{eq:R1_22c}\\[6pt]
{}_2R^{\times}_{2,2} &= \frac{t_1(5x_1^{4}+16x_1^{2}+4)}{16\,x_1}, \label{eq:R2_22c}\\[6pt]
{}_3R^{\times}_{2,2} &= \frac{4t_1x_1^{3}+\bar t_1\bar x_1(2x_1^{2}-1)}{8\,\bar t_1x_1}. \label{eq:R3_22c}
\end{align}
\end{tcolorbox}

\begin{tcolorbox}[breakable, colback=white, colframe=black!60, boxrule=0.5pt, arc=2pt,
  left=4pt, right=4pt, top=2pt, bottom=2pt, title={$\ell=3,\ m=0$}, coltitle=black, colbacktitle=black!8]
\begin{align}
{}_0R^{+}_{3,0} &= \frac{t_1}{15360\,\bar t_1x_1(\bar t_1^{2}+x_1^{2})^{5}}\Big[\bar t_1^{12}(6075x_1^{4}+5400x_1^{2}+720)+\bar t_1^{10}(30633x_1^{6}+16044x_1^{4}-9120x_1^{2}-1920) \nonumber\\
&\quad +\bar t_1^{8}x_1^{2}(61982x_1^{6}+280x_1^{4}-55680x_1^{2}-9600)+\bar t_1^{6}x_1^{4}(63426x_1^{6}-50544x_1^{4}-116640x_1^{2}-19200) \nonumber\\
&\quad +\bar t_1^{4}x_1^{6}(32295x_1^{6}-75760x_1^{4}-118480x_1^{2}-19200)+\bar t_1^{2}x_1^{8}(7061x_1^{6}-44092x_1^{4}-58880x_1^{2}-9600) \nonumber\\
&\quad -x_1^{10}(10048x_1^{4}+11968x_1^{2}+1920)\Big], \label{eq:R0_30p}\\[6pt]
{}_1R^{+}_{3,0} &= \frac{t_1\bar x_1(2x_1^{2}-1)[\bar t_1^{2}(2x_1^{2}+1)-x_1^{2}-1]}{8\,\bar t_1x_1^{2}}, \label{eq:R1_30p}\\[6pt]
{}_2R^{+}_{3,0} &= \frac{t_1[\bar t_1^{2}(405x_1^{4}+360x_1^{2}+48)-155x_1^{6}-508x_1^{4}-448x_1^{2}]}{1024\,x_1^{2}}, \label{eq:R2_30p}\\[6pt]
{}_3R^{+}_{3,0} &= \frac{4t_1x_1^{3}[\bar t_1^{2}(2x_1^{2}+1)-x_1^{2}-1]+\bar t_1\bar x_1[\bar t_1^{2}(4x_1^{4}-1)-4x_1^{4}-4x_1^{2}]}{16\,\bar t_1x_1^{2}}. \label{eq:R3_30p}
\end{align}
\end{tcolorbox}

\begin{tcolorbox}[breakable, colback=white, colframe=black!60, boxrule=0.5pt, arc=2pt,
  left=4pt, right=4pt, top=2pt, bottom=2pt, title={$\ell=3,\ m=1$}, coltitle=black, colbacktitle=black!8]
\begin{align}
{}_0R^{+}_{3,1} &= -\frac{\bar x_1}{46080\,\bar t_1x_1(\bar t_1^{2}+x_1^{2})^{5}}\Big[\bar t_1^{12}(12825x_1^{4}+12600x_1^{2}-720)+\bar t_1^{10}(63147x_1^{6}+49524x_1^{4}-37920x_1^{2}-5760) \nonumber\\
&\quad +\bar t_1^{8}x_1^{2}(124314x_1^{6}+61160x_1^{4}-176640x_1^{2}-28800)+\bar t_1^{6}x_1^{4}(123830x_1^{6}+4432x_1^{4}-340320x_1^{2}-57600) \nonumber\\
&\quad +\bar t_1^{4}x_1^{6}(59725x_1^{6}-53600x_1^{4}-331440x_1^{2}-57600)+\bar t_1^{2}x_1^{8}(12191x_1^{6}-39236x_1^{4}-160320x_1^{2}-28800) \nonumber\\
&\quad -x_1^{10}(10048x_1^{4}+32064x_1^{2}+5760)\Big], \label{eq:R0_31p}\\[6pt]
{}_1R^{+}_{3,1} &= -\frac{\bar x_1^{2}(2x_1^{2}-1)[\bar t_1^{2}(2x_1^{2}+3)-x_1^{2}-3]}{24\,\bar t_1x_1^{2}}, \label{eq:R1_31p}\\[6pt]
{}_2R^{+}_{3,1} &= -\frac{\bar x_1[\bar t_1^{2}(855x_1^{4}+840x_1^{2}-48)+15x_1^{6}-1028x_1^{4}-2560x_1^{2}-384]}{3072\,x_1^{2}}, \label{eq:R2_31p}\\[6pt]
{}_3R^{+}_{3,1} &= -\frac{4x_1^{3}\bar x_1[\bar t_1^{2}(2x_1^{2}+3)-x_1^{2}-3]+t_1\bar t_1\bar x_1^{2}(2x_1^{2}-1)(2x_1^{2}+3)}{48\,\bar t_1x_1^{2}}, \label{eq:R3_31p}\\[6pt]
{}_0R^{\times}_{3,1} &= -\frac{t_1\bar x_1}{2304\,\bar t_1(\bar t_1^{2}+x_1^{2})^{4}}\Big[\bar t_1^{10}(963x_1^{2}+396)+\bar t_1^{8}(3924x_1^{4}+534x_1^{2}-192)+\bar t_1^{6}x_1^{2}(6038x_1^{4}-1746x_1^{2}-768) \nonumber\\
&\quad +\bar t_1^{4}x_1^{4}(4124x_1^{4}-4518x_1^{2}-1152)+\bar t_1^{2}x_1^{6}(1143x_1^{4}-3562x_1^{2}-768)-x_1^{8}(1024x_1^{2}+192)\Big], \label{eq:R0_31c}\\[6pt]
{}_1R^{\times}_{3,1} &= -\frac{t_1\bar x_1^{2}(2x_1^{2}-1)(3\bar t_1^{2}-1)}{12\,\bar t_1x_1}, \label{eq:R1_31c}\\[6pt]
{}_2R^{\times}_{3,1} &= -\frac{t_1\bar x_1[\bar t_1^{2}(321x_1^{2}+132)-61x_1^{4}-210x_1^{2}-160]}{768\,x_1}, \label{eq:R2_31c}\\[6pt]
{}_3R^{\times}_{3,1} &= -\frac{8t_1x_1^{3}\bar x_1(3\bar t_1^{2}-1)+\bar t_1\bar x_1^{2}(2x_1^{2}-1)(6\bar t_1^{2}-5)}{48\,\bar t_1x_1}. \label{eq:R3_31c}
\end{align}
\end{tcolorbox}

\begin{tcolorbox}[breakable, colback=white, colframe=black!60, boxrule=0.5pt, arc=2pt,
  left=4pt, right=4pt, top=2pt, bottom=2pt, title={$\ell=3,\ m=2$}, coltitle=black, colbacktitle=black!8]
\begin{align}
{}_0R^{+}_{3,2} &= \frac{t_1}{46080\,\bar t_1x_1(\bar t_1^{2}+x_1^{2})^{5}}\Big[\bar t_1^{12}(21375x_1^{4}+19800x_1^{2}+3600) \nonumber\\
&\quad +\bar t_1^{10}(109605x_1^{6}+114780x_1^{4}+52320x_1^{2}+5760)+\bar t_1^{8}x_1^{2}(225830x_1^{6}+273400x_1^{4}+205440x_1^{2}+28800) \nonumber\\
&\quad +\bar t_1^{6}x_1^{4}(235738x_1^{6}+340496x_1^{4}+369120x_1^{2}+57600) \nonumber\\
&\quad +\bar t_1^{4}x_1^{6}(127835x_1^{6}+233840x_1^{4}+345840x_1^{2}+57600)+\bar t_1^{2}x_1^{8}(27393x_1^{6}+77204x_1^{4}+163200x_1^{2}+28800) \nonumber\\
&\quad +x_1^{10}(10048x_1^{4}+32064x_1^{2}+5760)\Big], \label{eq:R0_32p}\\[6pt]
{}_1R^{+}_{3,2} &= \frac{t_1\bar x_1(2x_1^{2}-1)[\bar t_1^{2}(6x_1^{2}+3)+x_1^{2}+3]}{24\,\bar t_1x_1^{2}}, \label{eq:R1_32p}\\[6pt]
{}_2R^{+}_{3,2} &= \frac{t_1[\bar t_1^{2}(1425x_1^{4}+1320x_1^{2}+240)+145x_1^{6}+980x_1^{4}-64x_1^{2}-768]}{3072\,x_1^{2}}, \label{eq:R2_32p}\\[6pt]
{}_3R^{+}_{3,2} &= \frac{4t_1x_1^{3}[\bar t_1^{2}(6x_1^{2}+3)+x_1^{2}+3]+\bar t_1\bar x_1[\bar t_1^{2}(12x_1^{4}-3)-4x_1^{4}+20x_1^{2}+6]}{48\,\bar t_1x_1^{2}}, \label{eq:R3_32p}\\[6pt]
{}_0R^{\times}_{3,2} &= \frac{1}{1152\,\bar t_1(\bar t_1^{2}+x_1^{2})^{4}}\Big[\bar t_1^{10}(666x_1^{4}+1143x_1^{2}+252)+\bar t_1^{8}(2634x_1^{6}+3972x_1^{4}-138x_1^{2}-192) \nonumber\\
&\quad +\bar t_1^{6}x_1^{2}(3894x_1^{6}+4574x_1^{4}-2994x_1^{2}-768)+\bar t_1^{4}x_1^{4}(2470x_1^{6}+1228x_1^{4}-5670x_1^{2}-1152) \nonumber\\
&\quad +\bar t_1^{2}x_1^{6}(640x_1^{6}-837x_1^{4}-4090x_1^{2}-768)-x_1^{8}(512x_1^{4}+1120x_1^{2}+192)\Big], \label{eq:R0_32c}\\[6pt]
{}_1R^{\times}_{3,2} &= \frac{\bar x_1(x_1^{2}+2)(2x_1^{2}-1)(2\bar t_1^{2}-1)}{12\,\bar t_1x_1}, \label{eq:R1_32c}\\[6pt]
{}_2R^{\times}_{3,2} &= \frac{\bar t_1^{2}(222x_1^{4}+381x_1^{2}+84)-169x_1^{4}-322x_1^{2}-128}{384\,x_1}, \label{eq:R2_32c}\\[6pt]
{}_3R^{\times}_{3,2} &= \frac{(x_1^{2}+2)[2x_1^{3}(2\bar t_1^{2}-1)+t_1\bar t_1\bar x_1(2x_1^{2}-1)]}{12\,\bar t_1x_1}. \label{eq:R3_32c}
\end{align}
\end{tcolorbox}

\begin{tcolorbox}[breakable, colback=white, colframe=black!60, boxrule=0.5pt, arc=2pt,
  left=4pt, right=4pt, top=2pt, bottom=2pt, title={$\ell=3,\ m=3$}, coltitle=black, colbacktitle=black!8]
\begin{align}
{}_0R^{+}_{3,3} &= -\frac{\bar x_1}{15360\,\bar t_1x_1(\bar t_1^{2}+x_1^{2})^{5}}\Big[\bar t_1^{12}(14925x_1^{4}+19800x_1^{2}+4080)+\bar t_1^{10}(72295x_1^{6}+92100x_1^{4}+29280x_1^{2}+1920) \nonumber\\
&\quad +\bar t_1^{8}x_1^{2}(139874x_1^{6}+162440x_1^{4}+84480x_1^{2}+9600)+\bar t_1^{6}x_1^{4}(137070x_1^{6}+124432x_1^{4}+126240x_1^{2}+19200) \nonumber\\
&\quad +\bar t_1^{4}x_1^{6}(70225x_1^{6}+22240x_1^{4}+104080x_1^{2}+19200)+\bar t_1^{2}x_1^{8}(14891x_1^{6}-23636x_1^{4}+44480x_1^{2}+9600) \nonumber\\
&\quad -x_1^{10}(10048x_1^{4}-8128x_1^{2}-1920)\Big], \label{eq:R0_33p}\\[6pt]
{}_1R^{+}_{3,3} &= -\frac{\bar x_1^{2}(2x_1^{2}-1)[\bar t_1^{2}(2x_1^{2}+3)-x_1^{2}+1]}{8\,\bar t_1x_1^{2}}, \label{eq:R1_33p}\\[6pt]
{}_2R^{+}_{3,3} &= -\frac{\bar x_1[\bar t_1^{2}(995x_1^{4}+1320x_1^{2}+272)-165x_1^{6}-20x_1^{4}+1280x_1^{2}-384]}{1024\,x_1^{2}}, \label{eq:R2_33p}\\[6pt]
{}_3R^{+}_{3,3} &= -\frac{4x_1^{3}\bar x_1[\bar t_1^{2}(2x_1^{2}+3)-x_1^{2}+1]+t_1\bar t_1\bar x_1^{2}(2x_1^{2}-1)(2x_1^{2}+3)}{16\,\bar t_1x_1^{2}}, \label{eq:R3_33p}\\[6pt]
{}_0R^{\times}_{3,3} &= -\frac{t_1\bar x_1}{768\,\bar t_1(\bar t_1^{2}+x_1^{2})^{4}}\Big[\bar t_1^{10}(405x_1^{2}+180)+\bar t_1^{8}(1836x_1^{4}+1770x_1^{2}+192)+\bar t_1^{6}x_1^{2}(3194x_1^{4}+5202x_1^{2}+768) \nonumber\\
&\quad +\bar t_1^{4}x_1^{4}(2532x_1^{4}+6822x_1^{2}+1152)+\bar t_1^{2}x_1^{6}(673x_1^{4}+4138x_1^{2}+768)+x_1^{8}(1024x_1^{2}+192)\Big], \label{eq:R0_33c}\\[6pt]
{}_1R^{\times}_{3,3} &= -\frac{t_1\bar x_1^{2}(2x_1^{2}-1)(\bar t_1^{2}+1)}{4\,\bar t_1x_1}, \label{eq:R1_33c}\\[6pt]
{}_2R^{\times}_{3,3} &= -\frac{t_1\bar x_1[\bar t_1^{2}(135x_1^{2}+60)+53x_1^{4}+402x_1^{2}+32]}{256\,x_1}, \label{eq:R2_33c}\\[6pt]
{}_3R^{\times}_{3,3} &= -\frac{8t_1x_1^{3}\bar x_1(\bar t_1^{2}+1)+\bar t_1\bar x_1^{2}(2x_1^{2}-1)(2\bar t_1^{2}+1)}{16\,\bar t_1x_1}. \label{eq:R3_33c}
\end{align}
\end{tcolorbox}